\documentclass[aps,prl,twocolumn,groupedaddress]{revtex4}

\usepackage{color}
\definecolor{darkblue}{rgb}{0.0,0.0,1.0}
\definecolor{lightblue}{rgb}{0,0.7,1.0}
\usepackage{graphicx}
\usepackage{amsmath}
\usepackage{float}
\usepackage{overpic}

\newcommand{\mus}{\mu_{\mathrm{s}}}
\newcommand{\mub}{\mu_{\mathrm{B}}}
\newcommand{\kb}{k_{\mathrm{B}}}
\newcommand{\FDW}{F_{\mathrm{DW}}}
\newcommand{\apy}{\alpha_{\mathrm{Py}}}
\newcommand{\bt}{\beta_{\mathrm{t}}}
\newcommand{\bv}{\beta_{\mathrm{v}}}
\newcommand{\bna}{\beta^{\mathrm{n}}}
\newcommand{\bsr}{\beta^{\mathrm{s}}}

\begin{document}

\title{Non-adiabatic spin torque investigated using thermally activated magnetic domain wall dynamics}

\author{M.~Eltschka}
\affiliation{Fachbereich Physik, Universit\"at Konstanz, Universit\"atsstra\ss e 10, 78457 Konstanz, Germany}

\author{M.~W\"otzel}
\altaffiliation{also at: Center for Electron Nanoscopy, Technical University of Denmark, 2800 Kgs. Lyngby, Denmark}
\affiliation{Fachbereich Physik, Universit\"at Konstanz, Universit\"atsstra\ss e 10, 78457 Konstanz, Germany}

\author{T.~Kasama}
\affiliation{Center for Electron Nanoscopy, Technical University of Denmark, 2800 Kgs. Lyngby, Denmark}

\author{J.~Rhensius}
\altaffiliation{also at: Laboratory for Micro- and Nanotechnology, Paul Scherrer Institut, 5232 Villigen PSI, Switzerland}
\affiliation{Fachbereich Physik, Universit\"at Konstanz, Universit\"atsstra\ss e 10, 78457 Konstanz, Germany}

\author{S.~Krzyk}
\affiliation{Fachbereich Physik, Universit\"at Konstanz, Universit\"atsstra\ss e 10, 78457 Konstanz, Germany}

\author{U.~Nowak}
\affiliation{Fachbereich Physik, Universit\"at Konstanz, Universit\"atsstra\ss e 10, 78457 Konstanz, Germany}

\author{L.~J.~Heyderman}
\affiliation{Laboratory for Micro- and Nanotechnology, Paul Scherrer Institut, 5232 Villigen PSI, Switzerland}

\author{R.~E.~Dunin-Borkowski}
\affiliation{Center for Electron Nanoscopy, Technical University of Denmark, 2800 Kgs. Lyngby, Denmark}

\author{M.~Kl\"aui}
\email[]{mathias@klaeui.de}
\altaffiliation{also at: Laboratory of Nanomagnetism and Spin Dynamics, Ecole Polytechnique F\'ed\'erale de Lausanne (EPFL), 1015 Lausanne, Switzerland}

\altaffiliation{SwissFEL, Paul Scherrer Institut, 5232 Villigen PSI, Switzerland}
\affiliation{Fachbereich Physik, Universit\"at Konstanz, Universit\"atsstra\ss e 10, 78457 Konstanz, Germany}

\author{H.~J.~van Driel}
\affiliation{Institute for Theoretical Physics, Utrecht University, Leuvenlaan 4, 3584 CE Utrecht, The Netherlands}

\author{R.~A.~Duine}
\affiliation{Institute for Theoretical Physics, Utrecht University, Leuvenlaan 4, 3584 CE Utrecht, The Netherlands}

\date{\today}

\begin{abstract}
Using transmission electron microscopy, we investigate the thermally activated motion of domain walls (DWs) between two positions in permalloy ($\mathrm{Ni}_{80}\mathrm{Fe}_{20}$) nanowires at room temperature. We show that this purely thermal motion is well described by an Arrhenius law, allowing for a description of the DW as a quasi-particle in a 1D potential landscape. By injecting small currents, the potential is modified, allowing for the determination of the non-adiabatic spin torque: $\bt = 0.010\pm0.004$ for a transverse DW and $\bv =0.073\pm0.026$ for a vortex DW. The larger value is attributed to the higher magnetization gradients present.  
\end{abstract}


\maketitle

The controlled motion of magnetic domain walls (DWs) by the injection of spin-polarized currents \cite{yamaguchi2004, klaeui2005} has become an exciting field of research, since the detailed understanding of the interplay between the local magnetization and the spin angular momentum of the electron current is of scientific interest and essential for proposed applications \cite{parkin2008}. Thermal effects, resulting from Joule heating due to the high injected current densities necessary for DW displacement, have often been neglected and have in general been considered as detrimental in the past. However, when properly analyzed, thermally activated processes at temperatures even well below the Curie temperature can be used to provide a better understanding of the underlying physical mechanisms \cite{haenggi1990}. Current-induced DW displacement has been demonstrated successfully in experiments \cite{yamaguchi2004, klaeui2005, parkin2008}, but the underlying theory describing the torque exerted by spin-polarized conduction electrons on the local magnetization and, in particular, their quantitative values are still subject to controversial discussions \cite{thiaville2005, zhang2004, tatara2008, xiao2006}.

In order to account for the effect of the current, two spin torque terms have been added to the Landau-Lifshitz or Gilbert equation the latter of which reads for the current flowing along the $x$-axis \cite{thiaville2005, zhang2004, schieback2007}:
\begin{align}\label{eq_torque}
	\frac{\partial \mathbf{S}_i}{\partial t}= &-\frac{\gamma}{\mus}\mathbf{S}_i \times \mathbf{H}_i(t) + \alpha \mathbf{S}_i \times \frac{\partial \mathbf{S}_i}{\partial t}\nonumber \\ 
	&- u_x \frac{\partial \mathbf{S}_i}{\partial x} + \beta u_x \mathbf{S}_i \times \frac{\partial \mathbf{S}_i}{\partial x}
\end{align}
where $\mathbf{H}_i(t)$ are the effective fields, $\gamma=g\mub/\hbar$ is the gyromagnetic ratio, $\mathbf{S}_i = \boldsymbol{\mu}_i / \mu_{\mathrm{s}} $ represents the magnetic moment of unit length with $\mu_{\mathrm{s}} = |\boldsymbol{\mu}_i|$, $u_x=jPg\mu _\mathrm{B}/(2eM_{\mathrm{s}})$ is an effective velocity for current density $j$, $P$ is the polarization and $M_{\mathrm{s}}$ is the saturation magnetization \cite{schieback2007}. Gilbert damping is described by the constant $\alpha$ and the non-adiabaticity is characterized by the parameter $\beta$. While the third term (adiabatic spin torque) is well understood, the physical contributions to the non-adiabatic torque are still the subject of scientific debate, and both spin-flip scattering and linear momentum transfer due to non-adiabatic transport have been discussed \cite{zhang2004, tatara2008, xiao2006}. In particular, this torque becomes in a 1D model an effective force \cite{tatara2008}. While the purely adiabatic contribution is expected to prevail for wide DW spin structures, the non-adiabatic term has been proposed to be more effective in narrow DWs or vortices where high magnetization gradients occur \cite{xiao2006}. However, the relative magnitude of the non-adiabatic term and its relation to the damping constant and to the characteristic transport length scale, that the DW width has to be compared to, are still debated and depend on the model used \cite{xiao2006, tatara2008, stiles2007, smith2008, barnes2005}.

Different methods to determine $\beta$ have been put forward with partly contradicting results. One problem has been the influence of thermal effects, that stem from the Joule heating due to the high current densities that are required for DW movement and which can affect the physical mechanisms involved \cite{schieback2009}. However, a first step to analyze the spin torque effect using thermal effects was recently taken, when the thermally activated depinning processes induced by external fields, with simultaneous current injection, were studied in hard magnetic materials \cite{burrowes2009}. DWs are pinned at artificially structured notches in wires, defects intrinsic to the material or defects caused by fabrication, all of which can generate attractive potential wells \cite{klaeui2005, burrowes2009, klaeui2008, bryan2004, compton2010}. Due to these strong pinning sites in hard magnetic materials \cite{burrowes2009}, DWs can only be depinned by additional external magnetic fields deforming DWs prior to depinning. The resulting complicated spin structures make reliable conclusions about spin torque difficult without direct imaging, which has not been carried out so far.

\begin{figure}
	\includegraphics[width=8cm]{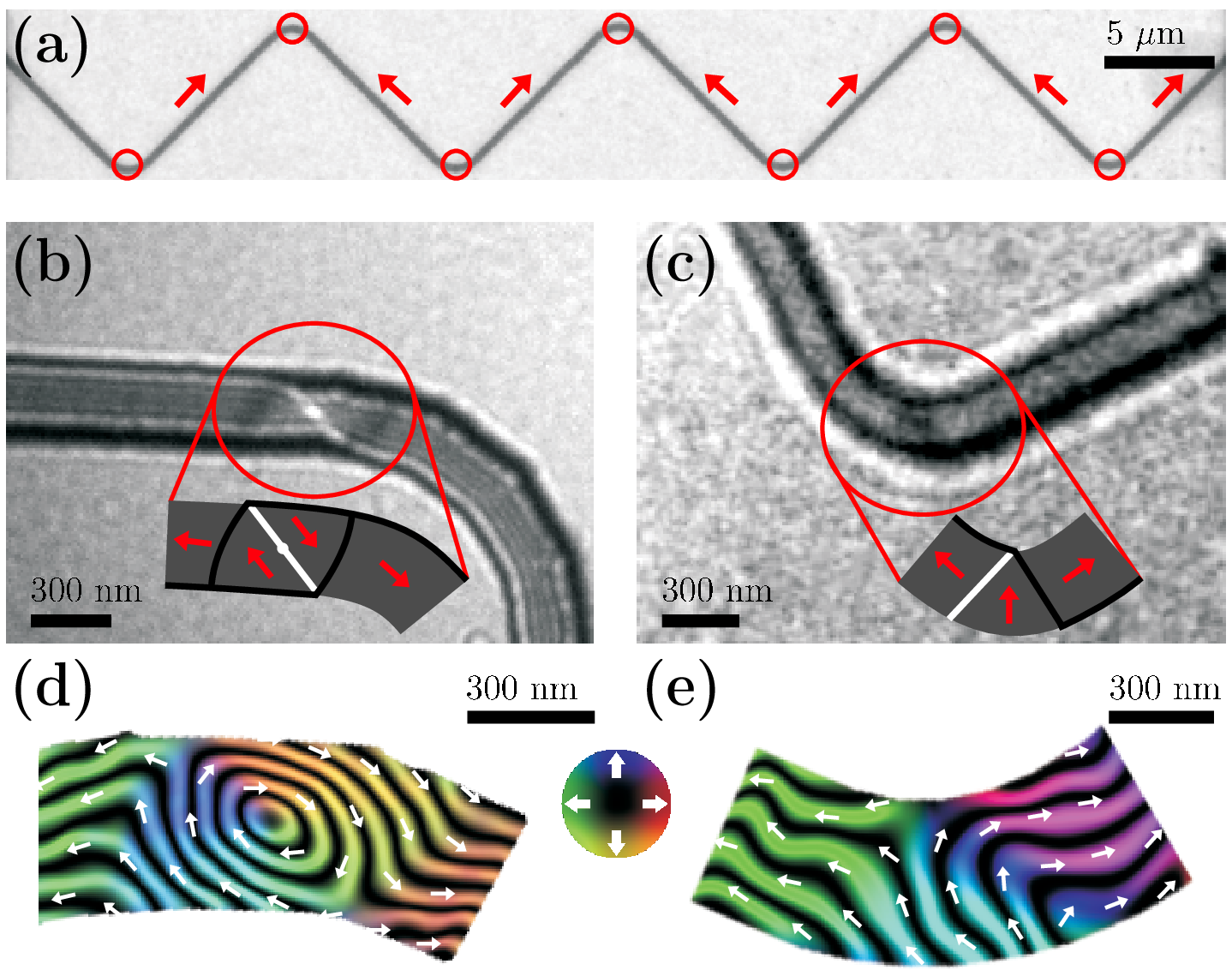}
	\caption{
	(a) Bright field TEM image of a permalloy zigzag line structure. The arrows indicate the directions of magnetization and the positions of nucleated DWs are marked by circles.
	(b) and (c) Fresnel images of thermally activated vortex and transverse DW jumping between two pinning sites.
	(d) and (e) Electron holography magnetic induction maps of vortex and transverse DW in the same wires showing detailed spin structures indicated by the color code and the arrows.}
	\label{fig_struc}
\end{figure}

Thus, in order to use thermal effects to analyze and fully understand the mechanisms behind spin torque, soft magnetic materials with weak DW pinning are necessary so that thermal effects can be observed at room temperature without the application of external fields. Only then can very small currents sufficiently affect the magnetic system without significant heating, so that the size of the spin torque terms [Eq.~\ref{eq_torque}] can be determined using direct time-resolved high resolution imaging of the spin structure and comparison to a 1D analytical description of thermally induced DW dynamics \cite{tatara2008, lucassen2009}. However, for DWs in soft magnetic materials such as permalloy, in which DWs are complicated 3D objects (see vortex DW spin structure [Fig.~\ref{fig_struc}(d)] and transverse DW spin structure [Fig.~\ref{fig_struc}(e)]), it is necessary to clarify if the description based on the quasi-particle model can be used and if thermal depinning occurs via a single path, which is necessary for an analysis based on a 1D model to determine the non-adiabatic spin torque.

In this paper, we investigate in real time thermally activated DWs jumping between two positions in permalloy ($\mathrm{Ni}_{80}\mathrm{Fe}_{20}$) nanowires using transmission electron microscopy. This motion occurs at room temperature and is of purely thermal origin, in the absence of an external magnetic field or injected electron current. The distribution of the dwell times for which a DW stays at each of the two positions is well described by an Arrhenius law with the DW described as a quasi-particle in a 1D potential, with two metastable states separated by an energy barrier. By modifying the local potential using a small constant current with no significant heating and by analyzing the dwell times, we infer values for $\beta$ for a transverse and a vortex DW. We find that there are significant differences between the values determined for $\beta$ for different DW types, which can be attributed to their different spin structures.
 
Fig.~\ref{fig_struc}(a) shows a transmission electron microscopy image of a permalloy zigzag wire on a silicon nitride membrane. These structures (wire width: 150-500 nm, wire thickness: 8-20 nm) were fabricated by electron-beam lithography and a lift-off procedure \cite{backes2006}. At both ends, the line structures are contacted by Au pads, allowing for the injection of currents. To improve heat dissipation, the membrane substrates can be additionally back coated with 30 nm of Aluminum. 

Transmission electron microscopy is a particularly suitable technique to observe and understand the full thermally activated stochastic dynamics of DWs as it provides high spatial ($<10$ nm) and time resolution ($<0.05$ s) for single shot measurements. Here, the Fresnel mode of Lorentz microscopy is used to identify the positions of the DWs from the dark or bright contrast that results from the deflection of electrons by the Lorentz force \cite{chapman1989}. For the investigation of detailed DW spin structures, we use off-axis electron holography \cite{borkowski2004}.

The sample is initially magnetized using a field of $\sim1$ T in a direction perpendicular to the wires. Tilting the sample by 30$^\circ$ results in the nucleation of DWs in the kinks of the wires after relaxing the field (see schematic in Fig.~\ref{fig_struc}(a)). We find either vortex [Fig.~\ref{fig_struc}(b) and (d)] or transverse [Fig.~\ref{fig_struc}(c) and (e)] DWs depending on the wire geometry \cite{klaeui2008}.

We first analyze a transverse DW (sample 1: wire width 495 nm, wire thickness 10 nm) jumping between two pinning sites (separation $318\pm$32 nm) without the influence of an external magnetic field or current. This pure thermal DW motion occurs at room temperature since no significant heating effects due to the electron beam of the transmission electron microscope were observed (The sample temperature is monitored by measuring the resistance \cite{laufenberg2006b}). After acquiring 7500 Lorentz images of the DW (DW width: $590\pm10$ nm) in the movie mode, we identify the DW positions and investigate the dwell times $\tau_1$ and $\tau_2$ that the DW stays at each of the two positions. 

\begin{figure}
	\includegraphics[width=8cm]{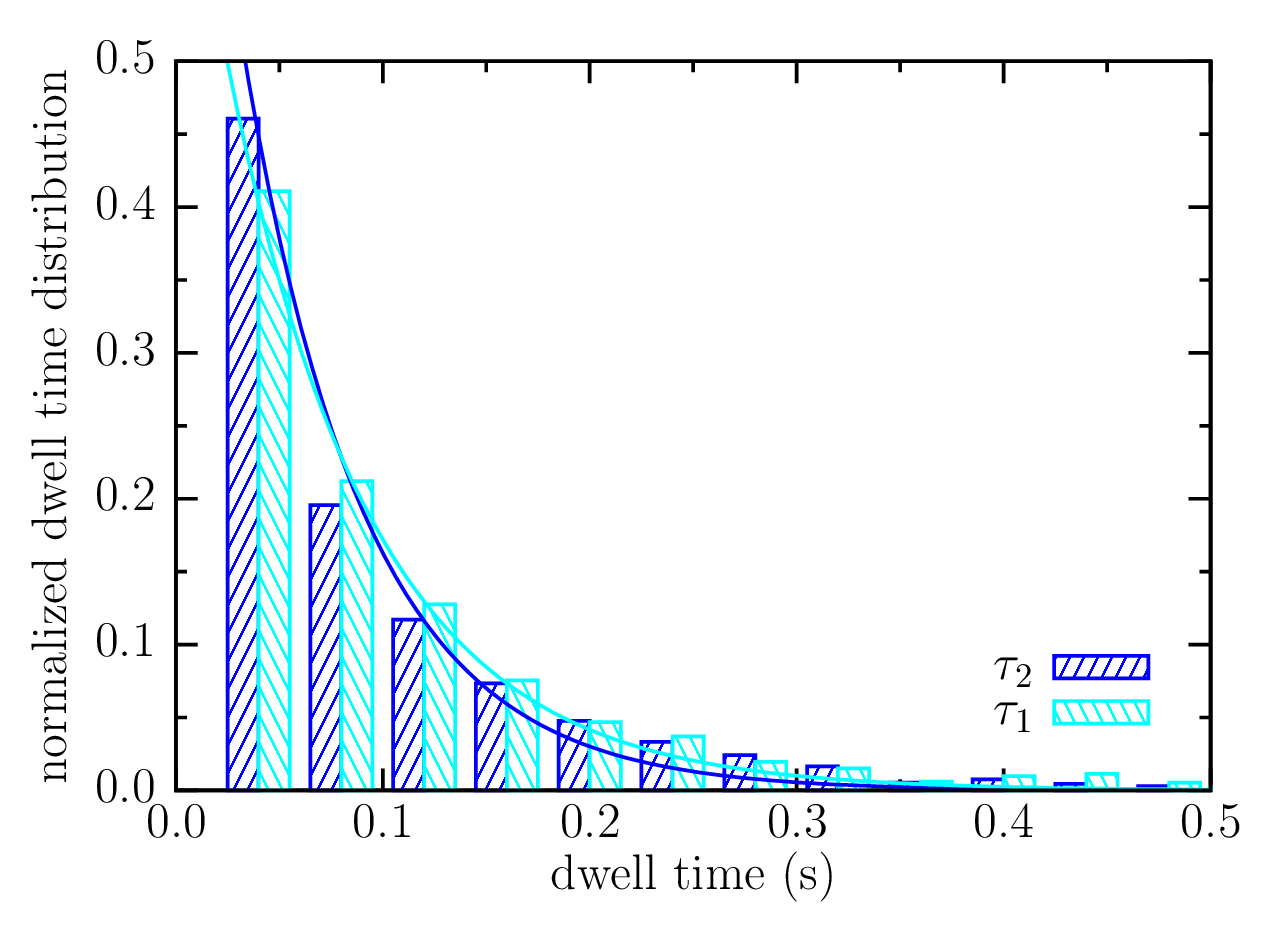}
	\caption{Normalized distribution of dwell times for both pinning sites ($\tau_1$ and $\tau_2$) of a thermally activated transverse DW in a permalloy wire (sample 1) fitted with a single exponential function.}
	\label{fig_tau}
\end{figure}

\begin{figure}
	\includegraphics[width=8cm]{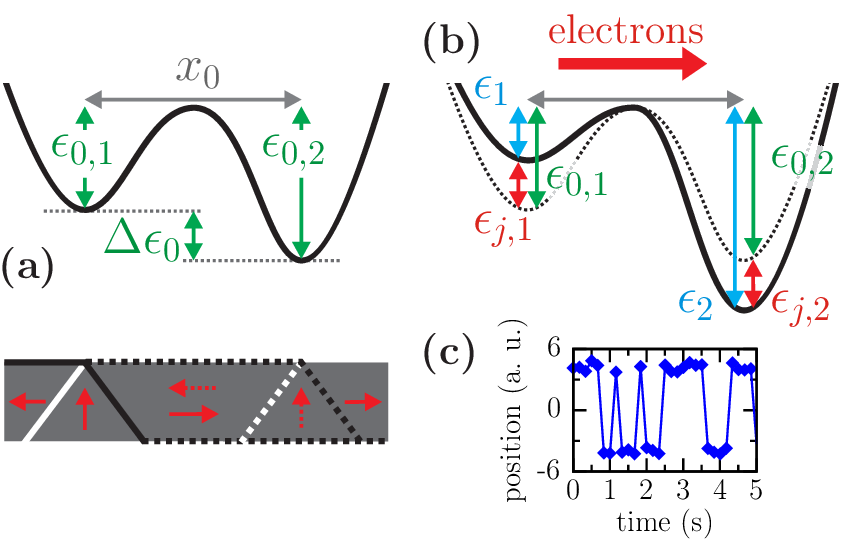}
	\caption{ 
	(a) Double well potential for a 1D description of thermally activated DWs jumping between two positions (schematically shown for a transverse DW). For the case without a current, the energy barriers consist of current independent contributions $\epsilon_{0,1}$ and $\epsilon_{0,2}$. The difference between the two barriers is $\Delta \epsilon_{0}$, the distance between the two pinning sites is $x_0$. 
	(b) An additional force on the DW results from the injection of a spin-polarized current due to the non-adiabatic spin torque. The energy barriers now consist of current independent $\epsilon_{0,i}$ and current dependent contributions $\epsilon_{j,i}$  ($i=1,~2$).
	(c) An example of measurements of the time-resolved thermal movement.}
	\label{fig_pot}
\end{figure}

The exponential decay of the dwell time distributions [Fig.~\ref{fig_tau}], and in particular the fact that it is well approximated by a single exponential function \cite{attane2006}, means that the DW can be considered as a quasi-particle moving in a 1D potential between two metastable states separated by an energy barrier, with a single transition path across it (shown schematically in Fig.~\ref{fig_pot}(a)) \cite{burrowes2009}. In general, the pinning strength is not equal at the two positions, yielding two depths for the potential wells with two different energy barriers $\epsilon_{0,1}$ and $\epsilon_{0,2}$. The probability of a transition to the neighboring state is only determined by the ratio between the corresponding energy barrier and the thermal energy that excites the transition. This dependence results in finite dwell times $\tau_1$ and $\tau_2$, for which the DW stays at each of the two local metastable states before it passes to the neighboring minimum of the pinning potential.

We carry out the same investigation for a vortex DW (sample 2: wire width 395 nm, wire thickness 17 nm). The two pinning positions (separation $20\pm4$ nm) of the DW (DW width $363\pm10$ nm) are again extracted from Lorentz [Fig.~\ref{fig_struc}(b)] and holography images [Fig.~\ref{fig_struc}(d)] and we find that the complete DW is displaced. The dwell times show the same exponential decay as observed above for the transverse DW suggesting that the DW transitions can again be described by an Arrhenius law.

Since the dwell times of both DW types obey an Arrhenius law, we use in the following a 1D quasi-particle description \cite{tatara2008} to analyze the influence of a constant direct current on the dynamics of the thermally activated DWs. In this model, the non-adiabatic torque acts as a force on the DW, $\FDW=-N \hbar \beta \tilde{u}_x/\lambda ^2$ where $N$ is the number of spins in the DW, $\tilde{u}_x$ is an effective spin current drift velocity and $\lambda$ is the DW width \cite{tatara2008, duine2007, lucassen2009b}. This force affects the 1D potential [Fig.~\ref{fig_pot}(b)] and the current dependent shift in the energy $\epsilon_j$ is:
\begin{eqnarray}\label{eq_1d1}
	\epsilon_j&=& \frac{N \hbar \beta \tilde{u}_x }{\lambda ^2} x  = \frac{2A\hbar\beta P}{e} \cdot \frac{j}{\lambda} \cdot x   
\end{eqnarray}
with $N=2\lambda A/a^3$ and $\tilde{u}_x=a^3Pj/e$, $A$ the cross-sectional area of the DW, $a^3$ the volume of the unit cell and the spin polarization ($P=0.37$ \cite{soulen1998}). The characteristic dwell times, for which the DW stays at each of the two positions are described by the Arrhenius law \cite{haenggi1990}:
\begin{eqnarray}\label{eq_1d2}
	\frac{\tau_1}{\tau_2}&=& \frac{\tau_{0,1}}{\tau_{0,2}} \cdot \mathrm{e}^{\left(\epsilon_{0,1} + \epsilon_{j,1}-\left(\epsilon_{0,2} + \epsilon_{j,2}\right) \right)/\left(\kb T\right)} 
\end{eqnarray}
Since the small current densities only induce a linear contribution to the current independent pinning potential (Eq.~\ref{eq_1d1}), the curvature of the potential at the two metastable states, which is given by the second derivative, is not significantly affected by the applied currents. Thus, the attempt frequencies $1/\tau_{0,1}$ and $1/\tau_{0,2}$ determined by the potential curvature \cite{haenggi1990} are independent of the current. From Eq.~\ref{eq_1d1} and Eq.~\ref{eq_1d2}, we obtain: 
\begin{eqnarray}
	\ln \frac{\tau_1}{\tau_2}  &=& \ln \frac{\tau_{0,1}}{\tau_{0,2}}  + \frac{\epsilon_{0,1} - \epsilon_{0,2}}{\kb T} + \frac{2A\hbar\beta P}{\kb T e} \cdot \frac{x_0}{\lambda} \cdot j \label{eq_fit}
\end{eqnarray}
By measuring dwell times as a function of current density and fitting these with Eq.~\ref{eq_fit}, we obtain values for $\beta$.
 
\begin{figure}
	\includegraphics[width=8cm]{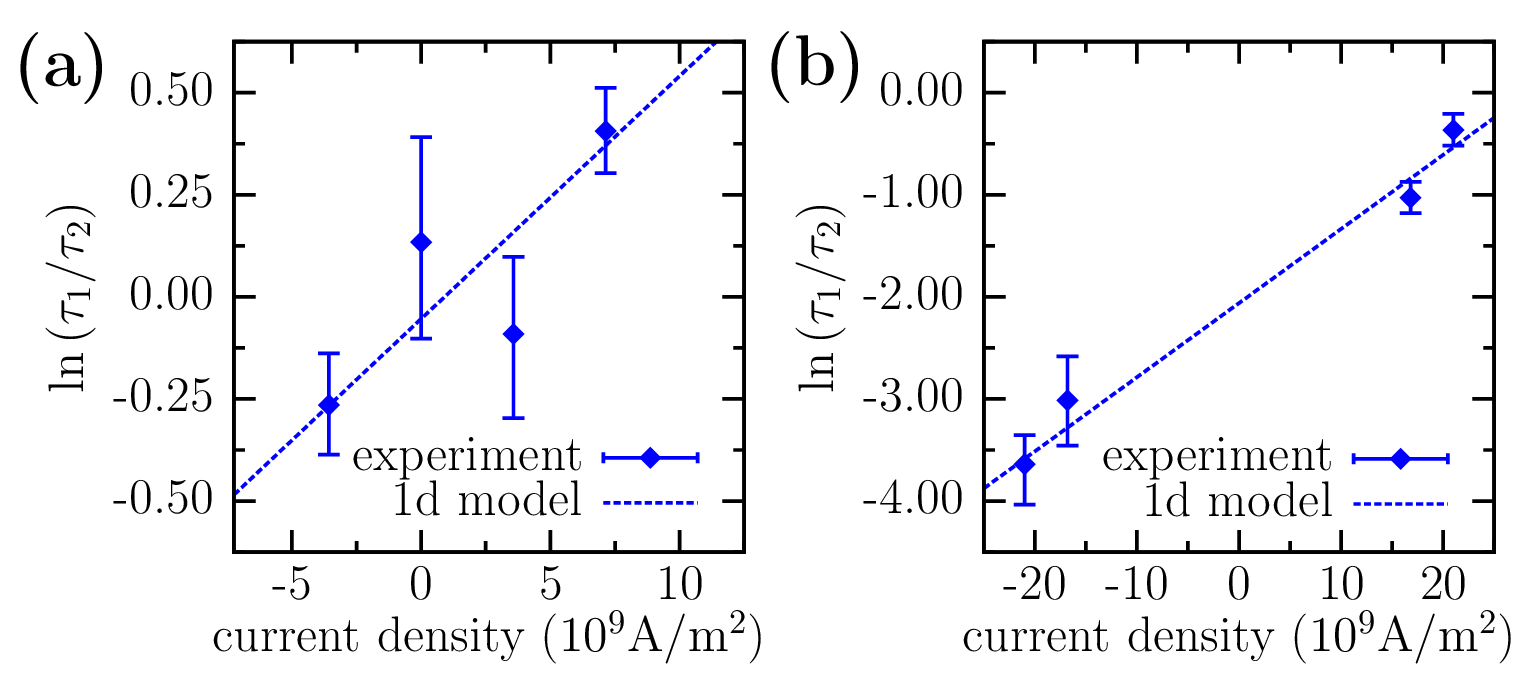}
	\caption{Measurements of dwell times $\tau_1$ and $\tau_2$ plotted as a function of applied direct current for (a) transverse and a (b) vortex DWs. The non-adiabatic coefficient $\beta$ is calculated from the slope of the fit with the 1D model (Eq.~\ref{eq_fit}).}
	\label{fig_beta}
\end{figure}

For a transverse DW (sample 1), we measure the dependence of dwell times on the intensities and directions of constant direct currents. For each current density, 7500 Lorentz images are acquired in order to obtain sufficient statistics when extracting the positions of the DW (see example in Fig.~\ref{fig_pot}(c)). The resultant dwell times for both metastable states [Fig.~\ref{fig_beta}(a)] are fitted to the 1D model (Eq.~\ref{eq_fit}) in order to calculate $\beta$ from the slope of the fit. From the Lorentz images and the electron holograms, values for the jump distance $x_0$ and the DW width $\lambda$ are derived and from the fit in Fig.~\ref{fig_beta}(a) we obtain $\bt = 0.010\pm0.004$ as result. This value is similar to the value of the Gilbert damping constant measured for our permalloy, $\apy \approx 0.008$ \cite{walowski2008}. 

For the vortex DW (sample 2), positions and dwell times are again extracted from Lorentz images, as shown in Fig.~\ref{fig_beta}(b), to obtain a value for $\bv =0.073\pm0.026$, which is considerably larger than the Gilbert damping constant. This measurement results in a larger non-adiabatic coefficient for the vortex DW than for the transverse DW ($\bt \approx 1.3 \cdot \apy$ for the transverse DW; $\bv \approx 9.2 \cdot \apy$ for the vortex DW). 

The fact that $\bv \gg  \bt$ can be explained by using a more sophisticated description of the non-adiabatic torque, which includes two contributions $\bv = \bna + \bsr$ with $\bna$ accounting for non-adiabatic transport and $\bsr$ for spin relaxation due to spin-flip scattering \cite{tatara2008a}. Since the magnetization inside a transverse DW varies slowly, the corresponding magnetization gradients are small, and non-adiabatic contributions to the spin torque effect are mainly the result of spin-flip scattering due to impurities and spin-orbit coupling \cite{tatara2008}. Therefore, $\bt \approx \bsr$ is expected for a transverse DW in permalloy and values of the order of the damping constant $\alpha$ are predicted \cite{barnes2005, tserkovnyak2006}. For a vortex DW in the investigated permalloy nanowires, the magnetization turns out-of-plane at the vortex core and a large magnetization gradient occurs. Thus, non-adiabatic transport due to reflected conduction electrons becomes more significant and $\bna$ is larger \cite{xiao2006},  while the contribution due to spin relaxation $\bsr$ is still present. Assuming that the contribution from spin relaxation $\bsr$ is similar for transverse and vortex DWs as it is intrinsic to the material, we can derive the pure contribution of non-adiabatic transport $\bna = \bv - \bt = 0.063\pm0.030$. We note that these results are deduced from analysis based on a 1D model, which nevertheless seems to be a reasonable assumption given the good fit to the Arrhenius law in Fig.~\ref{fig_tau}.

In conclusion, we derive $\beta$ from thermally activated vortex and a transverse DW obtaining $\bv = 0.073\pm0.026 \gg \bt = 0.010\pm0.004$. The difference between these values is attributed to the high magnetization gradients at the vortex core. The comparatively large value for $\bv$ means that high DW velocities ($>100$ m/s) should be possible for vortex walls, which bodes well for fast switching devices based on DW motion.  

We acknowledge support by the German Science Foundation (DFG SFB 767, KL 1811), EU (RTN Spinswitch MRTN-CT-2006-035327, Stg MASPIC ERC-2007-Stg 208162, Stg NEWSPIN ERC-2007-Stg 201350) and the Samsung Advanced Institute of Technology. M.~E.~thanks the German National Academic Foundation.


\begin{thebibliography}{29}
\expandafter\ifx\csname natexlab\endcsname\relax\def\natexlab#1{#1}\fi
\expandafter\ifx\csname bibnamefont\endcsname\relax
  \def\bibnamefont#1{#1}\fi
\expandafter\ifx\csname bibfnamefont\endcsname\relax
  \def\bibfnamefont#1{#1}\fi
\expandafter\ifx\csname citenamefont\endcsname\relax
  \def\citenamefont#1{#1}\fi
\expandafter\ifx\csname url\endcsname\relax
  \def\url#1{\texttt{#1}}\fi
\expandafter\ifx\csname urlprefix\endcsname\relax\def\urlprefix{URL }\fi
\providecommand{\bibinfo}[2]{#2}
\providecommand{\eprint}[2][]{\url{#2}}

\bibitem[{\citenamefont{Yamaguchi et~al.}(2004)\citenamefont{Yamaguchi, Ono,
  Nasu, Miyake, Mibu, and Shinjo}}]{yamaguchi2004}
\bibinfo{author}{\bibfnamefont{A.}~\bibnamefont{Yamaguchi}},
  \bibinfo{author}{\bibfnamefont{T.}~\bibnamefont{Ono}},
  \bibinfo{author}{\bibfnamefont{S.}~\bibnamefont{Nasu}},
  \bibinfo{author}{\bibfnamefont{K.}~\bibnamefont{Miyake}},
  \bibinfo{author}{\bibfnamefont{K.}~\bibnamefont{Mibu}}, \bibnamefont{and}
  \bibinfo{author}{\bibfnamefont{T.}~\bibnamefont{Shinjo}},
  \bibinfo{journal}{Phys. Rev. Lett.} \textbf{\bibinfo{volume}{92}},
  \bibinfo{pages}{077205} (\bibinfo{year}{2004}).

\bibitem[{\citenamefont{Kl\"aui et~al.}(2005)\citenamefont{Kl\"aui, Vaz, Bland,
  Wernsdorfer, Faini, Cambril, Heyderman, Nolting, and R\"udiger}}]{klaeui2005}
\bibinfo{author}{\bibfnamefont{M.}~\bibnamefont{Kl\"aui}},
  \bibinfo{author}{\bibfnamefont{C.~A.~F.} \bibnamefont{Vaz}},
  \bibinfo{author}{\bibfnamefont{J.~A.~C.} \bibnamefont{Bland}},
  \bibinfo{author}{\bibfnamefont{W.}~\bibnamefont{Wernsdorfer}},
  \bibinfo{author}{\bibfnamefont{G.}~\bibnamefont{Faini}},
  \bibinfo{author}{\bibfnamefont{E.}~\bibnamefont{Cambril}},
  \bibinfo{author}{\bibfnamefont{L.~J.} \bibnamefont{Heyderman}},
  \bibinfo{author}{\bibfnamefont{F.}~\bibnamefont{Nolting}}, \bibnamefont{and}
  \bibinfo{author}{\bibfnamefont{U.}~\bibnamefont{R\"udiger}},
  \bibinfo{journal}{Phys. Rev. Lett.} \textbf{\bibinfo{volume}{94}},
  \bibinfo{pages}{106601} (\bibinfo{year}{2005}).

\bibitem[{\citenamefont{Parkin et~al.}(2008)\citenamefont{Parkin, Hayashi, and
  Thomas}}]{parkin2008}
\bibinfo{author}{\bibfnamefont{S.~S.~P.} \bibnamefont{Parkin}},
  \bibinfo{author}{\bibfnamefont{M.}~\bibnamefont{Hayashi}}, \bibnamefont{and}
  \bibinfo{author}{\bibfnamefont{L.}~\bibnamefont{Thomas}},
  \bibinfo{journal}{Science} \textbf{\bibinfo{volume}{320}},
  \bibinfo{pages}{190} (\bibinfo{year}{2008}).

\bibitem[{\citenamefont{H\"anggi et~al.}(1990)\citenamefont{H\"anggi, Talkner,
  and Borkovec}}]{haenggi1990}
\bibinfo{author}{\bibfnamefont{P.}~\bibnamefont{H\"anggi}},
  \bibinfo{author}{\bibfnamefont{P.}~\bibnamefont{Talkner}}, \bibnamefont{and}
  \bibinfo{author}{\bibfnamefont{M.}~\bibnamefont{Borkovec}},
  \bibinfo{journal}{Rev. Mod. Phys} \textbf{\bibinfo{volume}{62}},
  \bibinfo{pages}{251} (\bibinfo{year}{1990}).

\bibitem[{\citenamefont{Thiaville et~al.}(2005)\citenamefont{Thiaville,
  Nakatani, Miltat, and Suzuki}}]{thiaville2005}
\bibinfo{author}{\bibfnamefont{A.}~\bibnamefont{Thiaville}},
  \bibinfo{author}{\bibfnamefont{Y.}~\bibnamefont{Nakatani}},
  \bibinfo{author}{\bibfnamefont{J.}~\bibnamefont{Miltat}}, \bibnamefont{and}
  \bibinfo{author}{\bibfnamefont{Y.}~\bibnamefont{Suzuki}},
  \bibinfo{journal}{Europhys. Lett.} \textbf{\bibinfo{volume}{69}},
  \bibinfo{pages}{990} (\bibinfo{year}{2005}).

\bibitem[{\citenamefont{Zhang and Li}(2004)}]{zhang2004}
\bibinfo{author}{\bibfnamefont{S.}~\bibnamefont{Zhang}} \bibnamefont{and}
  \bibinfo{author}{\bibfnamefont{Z.}~\bibnamefont{Li}}, \bibinfo{journal}{Phys.
  Rev. Lett.} \textbf{\bibinfo{volume}{93}}, \bibinfo{pages}{127204}
  (\bibinfo{year}{2004}).

\bibitem[{\citenamefont{Tatara et~al.}(2008)\citenamefont{Tatara, Kohno, and
  Shibata}}]{tatara2008}
\bibinfo{author}{\bibfnamefont{G.}~\bibnamefont{Tatara}},
  \bibinfo{author}{\bibfnamefont{H.}~\bibnamefont{Kohno}}, \bibnamefont{and}
  \bibinfo{author}{\bibfnamefont{J.}~\bibnamefont{Shibata}},
  \bibinfo{journal}{Phys. Rep.} \textbf{\bibinfo{volume}{468}},
  \bibinfo{pages}{213} (\bibinfo{year}{2008}).

\bibitem[{\citenamefont{Xiao et~al.}(2006)\citenamefont{Xiao, Zangwill, and
  Stiles}}]{xiao2006}
\bibinfo{author}{\bibfnamefont{J.}~\bibnamefont{Xiao}},
  \bibinfo{author}{\bibfnamefont{A.}~\bibnamefont{Zangwill}}, \bibnamefont{and}
  \bibinfo{author}{\bibfnamefont{M.~D.} \bibnamefont{Stiles}},
  \bibinfo{journal}{Phys. Rev. B} \textbf{\bibinfo{volume}{73}},
  \bibinfo{eid}{054428} (\bibinfo{year}{2006}).

\bibitem[{\citenamefont{Schieback et~al.}(2007)\citenamefont{Schieback,
  Kl\"aui, Nowak, R\"udiger, and Nielaba}}]{schieback2007}
\bibinfo{author}{\bibfnamefont{C.}~\bibnamefont{Schieback}},
  \bibinfo{author}{\bibfnamefont{M.}~\bibnamefont{Kl\"aui}},
  \bibinfo{author}{\bibfnamefont{U.}~\bibnamefont{Nowak}},
  \bibinfo{author}{\bibfnamefont{U.}~\bibnamefont{R\"udiger}},
  \bibnamefont{and} \bibinfo{author}{\bibfnamefont{P.}~\bibnamefont{Nielaba}},
  \bibinfo{journal}{Eur. Phys. J. B} \textbf{\bibinfo{volume}{59}},
  \bibinfo{pages}{429} (\bibinfo{year}{2007}).

\bibitem[{\citenamefont{Stiles et~al.}(2007)\citenamefont{Stiles, Saslow,
  Donahue, and Zangwill}}]{stiles2007}
\bibinfo{author}{\bibfnamefont{M.~D.} \bibnamefont{Stiles}},
  \bibinfo{author}{\bibfnamefont{W.~M.} \bibnamefont{Saslow}},
  \bibinfo{author}{\bibfnamefont{M.~J.} \bibnamefont{Donahue}},
  \bibnamefont{and} \bibinfo{author}{\bibfnamefont{A.}~\bibnamefont{Zangwill}},
  \bibinfo{journal}{Phys. Rev. B} \textbf{\bibinfo{volume}{75}},
  \bibinfo{eid}{214423} (\bibinfo{year}{2007}).

\bibitem[{\citenamefont{Smith}(2008)}]{smith2008}
\bibinfo{author}{\bibfnamefont{N.}~\bibnamefont{Smith}},
  \bibinfo{journal}{Phys. Rev. B} \textbf{\bibinfo{volume}{78}},
  \bibinfo{eid}{216401} (\bibinfo{year}{2008}).

\bibitem[{\citenamefont{Barnes and Maekawa}(2005)}]{barnes2005}
\bibinfo{author}{\bibfnamefont{S.~E.} \bibnamefont{Barnes}} \bibnamefont{and}
  \bibinfo{author}{\bibfnamefont{S.}~\bibnamefont{Maekawa}},
  \bibinfo{journal}{Phys. Rev. Lett.} \textbf{\bibinfo{volume}{95}},
  \bibinfo{pages}{107204} (\bibinfo{year}{2005}).

\bibitem[{\citenamefont{Schieback et~al.}(2009)\citenamefont{Schieback, Hinzke,
  Kl\"aui, Nowak, and Nielaba}}]{schieback2009}
\bibinfo{author}{\bibfnamefont{C.}~\bibnamefont{Schieback}},
  \bibinfo{author}{\bibfnamefont{D.}~\bibnamefont{Hinzke}},
  \bibinfo{author}{\bibfnamefont{M.}~\bibnamefont{Kl\"aui}},
  \bibinfo{author}{\bibfnamefont{U.}~\bibnamefont{Nowak}}, \bibnamefont{and}
  \bibinfo{author}{\bibfnamefont{P.}~\bibnamefont{Nielaba}},
  \bibinfo{journal}{Phys. Rev. B} \textbf{\bibinfo{volume}{80}},
  \bibinfo{pages}{214403} (\bibinfo{year}{2009}).

\bibitem[{\citenamefont{Burrowes et~al.}(2009)\citenamefont{Burrowes, Mihai,
  Ravelosona, Kim, Chappert, Vila, Marty, Samson, Garcia-Sanchez,
  Buda-Prejbeanu et~al.}}]{burrowes2009}
\bibinfo{author}{\bibfnamefont{C.}~\bibnamefont{Burrowes}},
  \bibinfo{author}{\bibfnamefont{A.~P.} \bibnamefont{Mihai}},
  \bibinfo{author}{\bibfnamefont{D.}~\bibnamefont{Ravelosona}},
  \bibinfo{author}{\bibfnamefont{J.-V.} \bibnamefont{Kim}},
  \bibinfo{author}{\bibfnamefont{C.}~\bibnamefont{Chappert}},
  \bibinfo{author}{\bibfnamefont{L.}~\bibnamefont{Vila}},
  \bibinfo{author}{\bibfnamefont{A.}~\bibnamefont{Marty}},
  \bibinfo{author}{\bibfnamefont{Y.}~\bibnamefont{Samson}},
  \bibinfo{author}{\bibfnamefont{F.}~\bibnamefont{Garcia-Sanchez}},
  \bibinfo{author}{\bibfnamefont{L.~D.} \bibnamefont{Buda-Prejbeanu}},
  \bibnamefont{et~al.}, \bibinfo{journal}{Nat. Phys.}
  \textbf{\bibinfo{volume}{6}}, \bibinfo{pages}{17} (\bibinfo{year}{2009}).

\bibitem[{\citenamefont{Kl\"{a}ui}(2008)}]{klaeui2008}
\bibinfo{author}{\bibfnamefont{M.}~\bibnamefont{Kl\"{a}ui}},
  \bibinfo{journal}{J. Phys.: Condens. Matter} \textbf{\bibinfo{volume}{20}},
  \bibinfo{pages}{313001} (\bibinfo{year}{2008}).

\bibitem[{\citenamefont{Bryan et~al.}(2004)\citenamefont{Bryan, Atkinson, and
  Cowburn}}]{bryan2004}
\bibinfo{author}{\bibfnamefont{M.~T.} \bibnamefont{Bryan}},
  \bibinfo{author}{\bibfnamefont{D.}~\bibnamefont{Atkinson}}, \bibnamefont{and}
  \bibinfo{author}{\bibfnamefont{R.~P.} \bibnamefont{Cowburn}},
  \bibinfo{journal}{Appl. Phys. Lett.} \textbf{\bibinfo{volume}{85}},
  \bibinfo{pages}{3510} (\bibinfo{year}{2004}).

\bibitem[{\citenamefont{Compton et~al.}(2010)\citenamefont{Compton, Chen, and
  Crowell}}]{compton2010}
\bibinfo{author}{\bibfnamefont{R.~L.} \bibnamefont{Compton}},
  \bibinfo{author}{\bibfnamefont{T.~Y.} \bibnamefont{Chen}}, \bibnamefont{and}
  \bibinfo{author}{\bibfnamefont{P.~A.} \bibnamefont{Crowell}},
  \bibinfo{journal}{Phys. Rev. B} \textbf{\bibinfo{volume}{81}},
  \bibinfo{pages}{144412} (\bibinfo{year}{2010}).

\bibitem[{\citenamefont{Lucassen and Duine}(2009)}]{lucassen2009}
\bibinfo{author}{\bibfnamefont{M.~E.} \bibnamefont{Lucassen}} \bibnamefont{and}
  \bibinfo{author}{\bibfnamefont{R.~A.} \bibnamefont{Duine}},
  \bibinfo{journal}{Phys. Rev. B} \textbf{\bibinfo{volume}{80}},
  \bibinfo{pages}{144421} (\bibinfo{year}{2009}).

\bibitem[{\citenamefont{Backes et~al.}(2006)\citenamefont{Backes, Heyderman,
  David, Sch\"aublin, Kl\"aui, Ehrke, R\"{u}diger, Vaz, Bland, Kasama
  et~al.}}]{backes2006}
\bibinfo{author}{\bibfnamefont{D.}~\bibnamefont{Backes}},
  \bibinfo{author}{\bibfnamefont{L.}~\bibnamefont{Heyderman}},
  \bibinfo{author}{\bibfnamefont{C.}~\bibnamefont{David}},
  \bibinfo{author}{\bibfnamefont{R.}~\bibnamefont{Sch\"aublin}},
  \bibinfo{author}{\bibfnamefont{M.}~\bibnamefont{Kl\"aui}},
  \bibinfo{author}{\bibfnamefont{H.}~\bibnamefont{Ehrke}},
  \bibinfo{author}{\bibfnamefont{U.}~\bibnamefont{R\"{u}diger}},
  \bibinfo{author}{\bibfnamefont{C.}~\bibnamefont{Vaz}},
  \bibinfo{author}{\bibfnamefont{J.}~\bibnamefont{Bland}},
  \bibinfo{author}{\bibfnamefont{T.}~\bibnamefont{Kasama}},
  \bibnamefont{et~al.}, \bibinfo{journal}{Microelectron. Eng.}
  \textbf{\bibinfo{volume}{83}}, \bibinfo{pages}{1726 } (\bibinfo{year}{2006}).

\bibitem[{\citenamefont{Chapman}(1989)}]{chapman1989}
\bibinfo{author}{\bibfnamefont{J.}~\bibnamefont{Chapman}},
  \bibinfo{journal}{Mater. Sci. Eng. B} \textbf{\bibinfo{volume}{3}},
  \bibinfo{pages}{355 } (\bibinfo{year}{1989}).

\bibitem[{\citenamefont{Dunin-Borkowski
  et~al.}(2004)\citenamefont{Dunin-Borkowski, McCartney, and
  Smith}}]{borkowski2004}
\bibinfo{author}{\bibfnamefont{R.~E.} \bibnamefont{Dunin-Borkowski}},
  \bibinfo{author}{\bibfnamefont{M.~R.} \bibnamefont{McCartney}},
  \bibnamefont{and} \bibinfo{author}{\bibfnamefont{D.~J.} \bibnamefont{Smith}},
  \emph{\bibinfo{title}{Encyclopedia of Nanoscience and Nanotechnology}}, vol.
  \bibinfo{volume}{3, pp. 41 to 100} (\bibinfo{publisher}{American Scientific
  Publishers, Stevenson Ranch, CA}, \bibinfo{year}{2004}).

\bibitem[{\citenamefont{Laufenberg et~al.}(2006)\citenamefont{Laufenberg,
  B\"uhrer, Bedau, Melchy, Kl\"aui, Vila, Faini, Vaz, Bland, and
  R\"udiger}}]{laufenberg2006b}
\bibinfo{author}{\bibfnamefont{M.}~\bibnamefont{Laufenberg}},
  \bibinfo{author}{\bibfnamefont{W.}~\bibnamefont{B\"uhrer}},
  \bibinfo{author}{\bibfnamefont{D.}~\bibnamefont{Bedau}},
  \bibinfo{author}{\bibfnamefont{P.-E.} \bibnamefont{Melchy}},
  \bibinfo{author}{\bibfnamefont{M.}~\bibnamefont{Kl\"aui}},
  \bibinfo{author}{\bibfnamefont{L.}~\bibnamefont{Vila}},
  \bibinfo{author}{\bibfnamefont{G.}~\bibnamefont{Faini}},
  \bibinfo{author}{\bibfnamefont{C.~A.~F.} \bibnamefont{Vaz}},
  \bibinfo{author}{\bibfnamefont{J.~A.~C.} \bibnamefont{Bland}},
  \bibnamefont{and}
  \bibinfo{author}{\bibfnamefont{U.}~\bibnamefont{R\"udiger}},
  \bibinfo{journal}{Phys. Rev. Lett.} \textbf{\bibinfo{volume}{97}},
  \bibinfo{pages}{046602} (\bibinfo{year}{2006}).

\bibitem[{\citenamefont{Attan\'{e} et~al.}(2006)\citenamefont{Attan\'{e},
  Ravelosona, Marty, Samson, and Chappert}}]{attane2006}
\bibinfo{author}{\bibfnamefont{J.~P.} \bibnamefont{Attan\'{e}}},
  \bibinfo{author}{\bibfnamefont{D.}~\bibnamefont{Ravelosona}},
  \bibinfo{author}{\bibfnamefont{A.}~\bibnamefont{Marty}},
  \bibinfo{author}{\bibfnamefont{Y.}~\bibnamefont{Samson}}, \bibnamefont{and}
  \bibinfo{author}{\bibfnamefont{C.}~\bibnamefont{Chappert}},
  \bibinfo{journal}{Phys. Rev. Lett.} \textbf{\bibinfo{volume}{96}},
  \bibinfo{pages}{147204} (\bibinfo{year}{2006}).

\bibitem[{\citenamefont{Duine et~al.}(2007)\citenamefont{Duine, N\'u\~nez, and
  MacDonald}}]{duine2007}
\bibinfo{author}{\bibfnamefont{R.~A.} \bibnamefont{Duine}},
  \bibinfo{author}{\bibfnamefont{A.~S.} \bibnamefont{N\'u\~nez}},
  \bibnamefont{and} \bibinfo{author}{\bibfnamefont{A.~H.}
  \bibnamefont{MacDonald}}, \bibinfo{journal}{Phys. Rev. Lett.}
  \textbf{\bibinfo{volume}{98}}, \bibinfo{pages}{056605}
  (\bibinfo{year}{2007}).

\bibitem[{\citenamefont{Lucassen et~al.}(2009)\citenamefont{Lucassen, van
  Driel, Smith, and Duine}}]{lucassen2009b}
\bibinfo{author}{\bibfnamefont{M.~E.} \bibnamefont{Lucassen}},
  \bibinfo{author}{\bibfnamefont{H.~J.} \bibnamefont{van Driel}},
  \bibinfo{author}{\bibfnamefont{C.~M.} \bibnamefont{Smith}}, \bibnamefont{and}
  \bibinfo{author}{\bibfnamefont{R.~A.} \bibnamefont{Duine}},
  \bibinfo{journal}{Phys. Rev. B} \textbf{\bibinfo{volume}{79}},
  \bibinfo{pages}{224411} (\bibinfo{year}{2009}).

\bibitem[{\citenamefont{Soulen et~al.}(1998)\citenamefont{Soulen, Byers,
  Osofsky, Nadgorny, Ambrose, Cheng, Broussard, Tanaka, Nowak, Moodera
  et~al.}}]{soulen1998}
\bibinfo{author}{\bibfnamefont{R.~J.} \bibnamefont{Soulen}},
  \bibinfo{author}{\bibfnamefont{J.~M.} \bibnamefont{Byers}},
  \bibinfo{author}{\bibfnamefont{M.~S.} \bibnamefont{Osofsky}},
  \bibinfo{author}{\bibfnamefont{B.}~\bibnamefont{Nadgorny}},
  \bibinfo{author}{\bibfnamefont{T.}~\bibnamefont{Ambrose}},
  \bibinfo{author}{\bibfnamefont{S.~F.} \bibnamefont{Cheng}},
  \bibinfo{author}{\bibfnamefont{P.~R.} \bibnamefont{Broussard}},
  \bibinfo{author}{\bibfnamefont{C.~T.} \bibnamefont{Tanaka}},
  \bibinfo{author}{\bibfnamefont{J.}~\bibnamefont{Nowak}},
  \bibinfo{author}{\bibfnamefont{J.~S.} \bibnamefont{Moodera}},
  \bibnamefont{et~al.}, \bibinfo{journal}{Science}
  \textbf{\bibinfo{volume}{282}}, \bibinfo{pages}{85} (\bibinfo{year}{1998}).

\bibitem[{\citenamefont{Walowski et~al.}(2008)\citenamefont{Walowski, M\"uller,
  Djordjevic, M\"unzenberg, Kl\"aui, Vaz, and Bland}}]{walowski2008}
\bibinfo{author}{\bibfnamefont{J.}~\bibnamefont{Walowski}},
  \bibinfo{author}{\bibfnamefont{G.}~\bibnamefont{M\"uller}},
  \bibinfo{author}{\bibfnamefont{M.}~\bibnamefont{Djordjevic}},
  \bibinfo{author}{\bibfnamefont{M.}~\bibnamefont{M\"unzenberg}},
  \bibinfo{author}{\bibfnamefont{M.}~\bibnamefont{Kl\"aui}},
  \bibinfo{author}{\bibfnamefont{C.~A.~F.} \bibnamefont{Vaz}},
  \bibnamefont{and} \bibinfo{author}{\bibfnamefont{J.~A.~C.}
  \bibnamefont{Bland}}, \bibinfo{journal}{Phys. Rev. Lett.}
  \textbf{\bibinfo{volume}{101}}, \bibinfo{pages}{237401}
  (\bibinfo{year}{2008}).

\bibitem[{\citenamefont{Tatara and Entel}(2008)}]{tatara2008a}
\bibinfo{author}{\bibfnamefont{G.}~\bibnamefont{Tatara}} \bibnamefont{and}
  \bibinfo{author}{\bibfnamefont{P.}~\bibnamefont{Entel}},
  \bibinfo{journal}{Phys. Rev. B} \textbf{\bibinfo{volume}{78}},
  \bibinfo{eid}{064429} (\bibinfo{year}{2008}).

\bibitem[{\citenamefont{Tserkovnyak et~al.}(2006)\citenamefont{Tserkovnyak,
  Skadsem, Brataas, and Bauer}}]{tserkovnyak2006}
\bibinfo{author}{\bibfnamefont{Y.}~\bibnamefont{Tserkovnyak}},
  \bibinfo{author}{\bibfnamefont{H.~J.} \bibnamefont{Skadsem}},
  \bibinfo{author}{\bibfnamefont{A.}~\bibnamefont{Brataas}}, \bibnamefont{and}
  \bibinfo{author}{\bibfnamefont{G.~E.~W.} \bibnamefont{Bauer}},
  \bibinfo{journal}{Phys. Rev. B} \textbf{\bibinfo{volume}{74}},
  \bibinfo{eid}{144405} (\bibinfo{year}{2006}).

\end{thebibliography}
\end{document}